\documentclass[journal=jctc,manuscript=article]{achemso}

\usepackage{xcolor}
\usepackage[utf8]{inputenc}
\usepackage[T1]{fontenc}
\usepackage{graphicx}  
\usepackage{dcolumn}   
\usepackage{bm}        
\usepackage{amssymb}   
\usepackage{framed}
\usepackage{amsmath}
\usepackage{multirow}
\usepackage{hhline}
\usepackage{color}
\usepackage{xcolor}
\definecolor{bluegreen}{rgb}{0,0.2,0.8}
\usepackage{subfigure,amsmath,verbatim,moreverb}
\usepackage{tabularx}
\usepackage{adjustbox}
\usepackage{lipsum}
\usepackage{longtable}
\usepackage{booktabs}
\usepackage{adjustbox}
\usepackage{hyperref}
\UseRawInputEncoding
\usepackage{graphicx}  
\usepackage{threeparttable} 
\usepackage{rotating}
\usepackage{geometry}
\usepackage{threeparttable} 

\usepackage{etoolbox}
\AtBeginEnvironment{align}{\setcounter{subeqn}{0}}
\newcounter{subeqn} %

\newcommand{\R}{\mathbf{r}}

\newcommand{\funcder}[2]{\frac{\delta #1}{\delta #2}}

\newcommand{\Eq}[1]{Eq.~(\ref{#1})}

\newcommand{\RRef}[1]{Ref. \citenum{#1}}
\newcommand{\Refs}[2]{Ref. \citenum{#1,#2}}
\newcommand{\Fig}[1]{Fig. \ref{#1}}
\newcommand{\Tab}[1]{Table \ref{#1}}

\newcommand{\braket}[2]{\ensuremath{\langle  #1 \vert #2  \rangle}}
\newcommand{\abraket}[2]{\ensuremath{\langle  #1 \vert\vert #2  \rangle}}

\usepackage{color}

\newcommand{\rev}[1]{#1}

\author{Aditi Singh}
\affiliation[]
{Institute of Physics, Faculty of Physics, Astronomy and Informatics,
Nicolaus Copernicus University, Grudziadzka 5, 87-100 Toru\'n, Poland}
\author{Eduardo Fabiano}
\affiliation[Unknown University]
{Istituto Nanoscienze-CNR, Via per Arnesano 16, I-73100 Lecce, Italy}
\alsoaffiliation[Second University]
{Center for Biomolecular Nanotechnologies @UNILE, 
Istituto Italiano di Tecnologia (IIT), Via Barsanti, 73010 Arnesano (LE), Italy}
\author{Szymon \'Smiga}
\affiliation[]
{Institute of Physics, Faculty of Physics, Astronomy and Informatics,
Nicolaus Copernicus University, Grudziadzka 5, 87-100 Toru\'n, Poland}

\email{szsmiga@fizyka.umk.pl}

\title[An \textsf{achemso} demo]
  {Understanding the core limitations of second-order correlation-based functionals through: functional, orbital, and eigenvalue-driven analysis}

\begin{document}


\begin{abstract}
Density Functional Theory has long struggled to obtain the exact exchange-correlational (XC) functional. Numerous approximations have been designed with the hope of achieving chemical accuracy. However, designing a functional involves numerous methodologies, which has a greater possibility for error accumulation if the functionals are poorly formulated. This study aims to investigate the performance and limitations of second-order correlation functionals within the framework of density functional theory. Specifically, we focus on three major classes of density functional approximations that incorporate second-order energy expressions: \textit{ab initio} (primarily G\"{o}rling-Levy) functionals, adiabatic connection models, and double-hybrid functionals. The principal objectives of this research are to evaluate the accuracy of second-order correlation functionals, to understand how the choice of reference orbitals and eigenvalues affects the performance of these functionals, to identify the intrinsic limitations of second-order energy expressions, especially when using arbitrary orbitals or non-canonical configurations, and propose strategies for improving their accuracy. By addressing these questions, we aim to provide deeper insights into the factors governing the accuracy of second-order correlation functionals, thereby guiding future functional development.
\end{abstract}

\section{I. Introduction}
For nearly about sixty years, Kohn-Sham\cite{Kohn_sham} (KS) density functional theory (KS-DFT) has played an important role in the theoretical description of chemical and solid-state systems\cite{mardirossian_rev}. Despite its unquestionable success, the most significant drawback is the need for the approximate treatment of the exchange-correlation (XC) effects within KS-DFT formalism, directly impacting the method's accuracy. Several classes of density functional approximation (DFA) have been proposed over the years, starting from the most simple local density approximation (LDA)\cite{Kohn_sham} (depending only on the density $\rho(\R)$) and ending on the most sophisticated KS orbital and eigenvalue dependent ones\cite{grimme2007doublehybrid, furche01, AngLiuTouJan-JCTC-11,furche:2008:rpa,OEP-RPA,dRPADH,grabowski:2002:OEPP2,mori-sanchez:2005:oeppt2,engel:2005:oeppt2,grabowski:2007:ccpt2,bartlett:2005:abinit2,OEPSOSszs,SOSa, THszs}. In the case of the latter, we can distinguish the large group of DFAs that utilize the second-order correlation energy expression in the XC formula (e.g., the G\"orling-Levy\cite{gorling:1994:OEP,ivanov:2003:P2} correlation energy expression at second order - GL2 and semi canonically transformed\cite{bartlett:2005:abinit2} - SC), i.e., second-order \textit{ab initio} DFT functionals\cite{grabowski:2002:OEPP2,bartlett:2005:abinit2,engel:2005:oeppt2,schweigert:2006:pt2,mori-sanchez:2005:oeppt2}, the ones constructed from adiabatic connection (AC) models (ACM) which interpolates between known high and low-density limits of the AC integrand \cite{isi,isierr,SeiPerLev-PRA-99,gorigiorgi09,SeiPerLev-PRA-99,liu09,ernzrehof99,teale10}, and 
the double-hybrid (DH) DFAs\cite{grimme2007doublehybrid,sharkas11,bremond11,bremond16}. 
Appendix A provides a short overview of these methods.

Second-order methods strongly improve DFT's ability to predict chemical properties. Nevertheless, they often have accuracy limitations due to error accumulation in functional, orbital, and eigenvalue-dependent calculations.
Moreover, to perform full self-consistent field (SCF) KS calculations with the aforementioned types of XC DFA ($E_{xc}^{DFA}[\{\phi_{p\sigma}\},\{\varepsilon_{p\sigma}\}]$), one needs to compute corresponding XC potential ($v_{xc}(\R)$) via functional derivative relations 
\begin{eqnarray}
v_{xc}(\R) & = & \funcder{E_{xc}^{DFA}[\{\phi_{p\sigma}\},\{\varepsilon_{p\sigma}\}]}{\rho(\R)}  \; .
\label{e1}
\end{eqnarray}
In the case of orbital and eigenvalue-dependent DFAs, this cannot be done directly as in the case of semi-local DFAs. Thus, one must employ the optimized effective potential (OEP) method\cite{talman:1976:OEP,ivanov:1999:OEP, kummel:2008:oep} to compute the corresponding XC potential to remain fully in the KS realm. We remark that the OEP procedure is a standard path of finding the XC potential within \textit{ab initio} DFT framework. The same procedure was recently employed in the context of ACMs\cite{SCF-ISI} and DH\cite{DHOEPSmiga2016, DHRSOEP} DFAs. We also remark, however, that the solution of the OEP equation is still not an easy task\cite{OEP_AQC, NikitasOEP}, especially at correlated second-order level\cite{grabowski:2002:OEPP2,bartlett:2005:abinit2,mori-sanchez:2005:oeppt2,grabowski14_2}. Thus, usually, one avoids, in routine DHs and ACMs DFA calculation, the troublesome OEP procedure by feeding $E_{xc}^{DFA}[\{\phi_{p\sigma}\},\{\varepsilon_{p\sigma}\}]$ with orbitals obtained from different computational methods. In the case of ACMs, many types of orbitals have been tested\cite{fabianoisi16, isigold, LUCISI, gISI2}, showing their significant impact on the quality of the final results.
In the case of DHs DFA\cite{DHrev}, the orbitals and eigenvalues are usually obtained using a generalized KS (GKS) scheme\cite{frit86,seidl1996generalized}, where the GL2 term is disregarded in the XC potential. This means that for DHs DFA, \Eq{e1} is not completely satisfied, and the GKS equations are solved at the
hybrid level. The influence of second-order term on the quality of orbitals and DH predictions has also been investigated employing the orbital optimization (OO)\cite{OODH} DH approach, showing large importance in some cases\cite{OODH, OOAS, OOQIDH}. We remark, however, that in the OO-DH approach, the full XC potential is non-local, and it is very different from the OEP realization of DH\cite{DHOEPSmiga2016, DHRSOEP}.  We also note that there exist few DH functionals\cite{XYG3, PhysRevLett.117.133002,xDH-PBE} where relation \Eq{e1} is fully decoupled, meaning that XC potential and functional used in KS-DFT calculations have different expressions not linked by \Eq{e1}. As in the case of ACM functionals, DHs performance strictly depends on the choice of input orbitals and eigenvalues, shown to some extent in \RRef{JMGL2withHF}. In some cases, this decoupling leads to a large improvement of the results\cite{XYG, XYG3, JMXYg3}. 

It is evident that the error cancellation effect plays a crucial role in the performance of second-order-based XC functionals. Thus, fully understanding these effects is essential for identifying margins of improvement of the various methods and determining how such improvements can be achieved.
This work seeks to advance the development of second-order-based  XC
functionals by leveraging a hierarchy of error-decomposition formulas. We begin by obtaining nearly exact (NeX) KS orbitals and eigenvalues from inverted coupled-cluster calculations to set a reference "gold standard." \rev{We mostly focus on small systems where second-order based functionals behave relatively well.}
  Then, we perform a thorough analysis by differentiating between orbital-driven (OD) and eigenvalue-driven (ED) errors, proposing a refined framework for assessing the performance of \textit{ab initio}, ACM, and DH functionals. The primary goal of this study is to develop a more nuanced understanding of the intrinsic limitations of the second-order energy expressions and the error cancellation mechanisms, aiming to improve the predictive power of second-order DFAs in total energy, binding energy, and reaction energy calculations.
\rev{Here, it is important to highlight that even though we investigate a KS-DFT perspective in this work, there exist other stances (e.g., such as perturbation theory from the HF reference\cite{Daas2020}) that could result in a substantially different error decomposition.} 

Thus, we aim to address three fundamental questions: 
\begin{enumerate}
\item What are the primary sources of error in second-order correlation functionals, and how do these errors manifest across different types of chemical systems?
\item How do orbital- and eigenvalue-driven errors contribute to the overall performance of second-order correlation functionals? Which of them plays a dominant role in different chemical environments?
\item What strategies can be developed to improve error cancellation mechanisms within these functionals, particularly in the context of ACMs?
\end{enumerate}
This investigation will, therefore, likely set the stage for further functional development by substantially enhancing the predictive capacity of such a class of DFAs.\\

The paper is organized as follows. We discuss our methodology and present the computational details in Sec. II. The results are discussed in Sec. III. We finish with a conclusion and future perspective. A brief overview of the \textit{ab inito} DFT, ACM, and DH functional approximations is given in Appendix A.

\section{II. Method}
\label{sec_method}
The performance of second-order-based XC 
functionals in predicting chemical properties is closely tied to the choice of reference orbitals and eigenvalues. In this study, we investigate the practical limitations of these functionals by systematically comparing different orbital references and error decomposition approaches. To disentangle the various sources of inaccuracy, we follow an approach similar to the one introduced in Refs. \cite{kim13,sim18,MDDFDE} and thus, the error in the energy of a functional can be written as
\begin{equation}
\Delta E[\tilde{\mathit{R}}] \equiv \tilde{E}[\tilde{\mathit{R}}] - E[\mathit{R}] = \Delta E_{FD}[\mathit{R}] + \Delta E_{RD}[\tilde{\mathit{R}}]\; .
\end{equation}
The tilde quantities are approximate ones, whereas those without tilde are NeX values. The symbol $\mathit{R}$ in the square parenthesis indicates that the energy is computed for a given approximate (or NeX) set of orbitals and eigenvalues; in the following, we will refer to this simply as the \emph{reference}. 
The functional-driven (FD) error is defined as
\begin{equation}
\Delta E_{FD}[\mathit{R}] \equiv \tilde{E}[\mathit{R}] -  E[\mathit{R}] \ .
\end{equation}
It measures the error contribution due to the functional approximation, irrespective 
of the reference used to feed the density approximation. Note that because in the 
FD error expression the same reference is employed both for $\tilde{E}$ and $E$ several energy contributions, such as the kinetic and Coulomb ones, cancel exactly; thus, we have $\Delta E_{FD}[\mathit{R}] \equiv \tilde{E}_{xc}[\mathit{R}] -  E_{xc}[\mathit{R}]$.
The reference-driven (RD) error is instead a measure of the effect of employing an approximate reference in the calculation in place of the NeX one. Thus, it is defined as
\begin{equation}
\Delta E_{RD}[\tilde{\mathit{R}}] \equiv \tilde{E}[\tilde{\mathit{R}}] - \tilde{E}[\mathit{R}]\   
\end{equation}
which,  
unlike the FD one, depends on the choice of the reference. \rev{For the conventional semi-local XC approximations, the above analysis corresponds to  the one reported in \RRef{kim13,sim18,MDDFDE}, with $\Delta E_{RD}$ being the density-driven error. }
Nonetheless, this study focuses on second-order correlation
functionals, which rely directly on orbitals and eigenvalues. Therefore, the density analysis is insufficient to comprehend the outcomes. Some references might lead to the same density (e.g., this is the case for any wave function calculation, which connects to Wu-Yang inverted \cite{Wy2003} analog) but yield very different values of the XC energy. Hence, we further partition the 
RD error as
\begin{equation}
\Delta E_{RD}[\tilde{\mathit{R}}] \equiv \Delta E_{OD}[\tilde{\phi}] + \Delta E_{ED}[\tilde{\phi},\tilde{\epsilon}]\ ,
\end{equation}
where $\tilde{\phi}$ denotes the set of orbitals and $\tilde{\epsilon}$ denotes the set of eigenvalues constituting the reference $\tilde{\mathit{R}}$. 
The orbital-driven (OD) error is then defined as
\begin{equation}
\Delta E_{OD}[\tilde{\phi}] \equiv \tilde{E}[\tilde{\phi},\epsilon] - \tilde{E}[\phi,\epsilon]
\end{equation}
and it measures the impact of using different orbitals. In turn, the eigenvalue-driven (ED) error is defined by the formula
\begin{equation}
\Delta E_{ED}[\tilde{\phi},\tilde{\epsilon}] \equiv \tilde{E}[\tilde{\phi},\tilde{\epsilon}] - \tilde{E}[\tilde{\phi},\epsilon]
\end{equation}
and it measures the effect of using references with different eigenvalue spectra. Because all terms in DFT energy are independent of the eigenvalues, with the possible exception of correlation, it is simple to see that indeed $\Delta E_{ED}[\tilde{\phi},\tilde{\epsilon}] = \tilde{E}_c[\tilde{\phi},\tilde{\epsilon}] - \tilde{E}_c[\tilde{\phi},\epsilon]$. In fact, it is also possible to show that
\begin{equation}
  \Delta E_{ED} [\tilde{\phi},\tilde{\epsilon}]= \frac{1}{4} \sum_{abij} \gamma^{ab}_{ij} \left \lvert \left \langle ij \| ab \right \rangle \right \rvert^2 
\end{equation}
with 
\begin{equation}
 \gamma^{ab}_{ij} = \frac{\left( D^{ab}_{ij} - \widetilde{D}^{ab}_{ij} \right)}{D^{ab}_{ij} \widetilde{D}^{ab}_{ij}}\ ,
\end{equation}
where $\widetilde{D}^{ab}_{ij}$ and $D^{ab}_{ij}$ are the second-order denominators corresponding to $\widetilde{\rho}$ and NeX densities, respectively.

\subsection{A. Computational details}

In this study we have considered a few representative examples of orbital-dependent second-order XC energy expression from each category, namely: i) OEP-GL2 \cite{grabowski:2002:OEPP2} and OEP2-SC \cite{bartlett:2005:abinit2} for \textit{ab inito} DFT; ii) ISI \cite{isi} and SPL \cite{SeiPerLev-PRA-99} from ACMs with the hPC \cite{SCF-ISI} model to evaluate the $W_\infty$ and $W'_\infty$ ingredients; iii) double hybrid functionals: empirical B2PLYP \cite{grimme2007doublehybrid}, non-empirical PBE-QIDH\cite{BApaper} and XYG3\cite{XYG3} as an example of fully decoupled relation \Eq{e1}. Here, we have also considered the recently developed BL1P functional\cite{bl1p} optimized for the density of HF \rev{(see Appendix A for more details)}.

To evaluate the impact of the employed reference, namely the orbitals and eigenvalues, on the performance of all the functionals, we have considered several possibilities: 
\begin{enumerate}
    \item Orbitals and eigenvalues from \emph{standard} approaches, that are HF, PBE\cite{pbe} and PBE0\cite{pbe0} methods.
    \item Self-consistent orbitals and eigenvalues from the KS OEP approach. \rev{For all technical details regarding the KS OEP realization of second-order functionals, we refer the reader to \RRef{DHOEPSmiga2016, SmigaJCP2020, SCF-ISI, OEP_AQC}}. 
    For double-hybrid functionals, we have also considered the orbitals and eigenvalues obtained from the GKS approach, which is the standard way these functionals are employed. 
    \item Orbitals and eigenvalues obtained from the direct optimization technique of Wu and Yang (WY) \cite{Wy2003}. In this case, we have considered several starting points, namely HF, second-order M{\o}ller-Plesset (MP2), and coupled-cluster singles-doubles (CCSD) relaxed density matrices.
\end{enumerate}
Tight convergence criteria were enforced for all SCF calculations, corresponding to maximum deviations in density matrix elements of $10^{-8}$ a.u. As NeX reference, we have utilized in all cases the orbitals and eigenvalues obtained using the WY method, taking the CCSD(T) \cite{Raghavachari1989479} relaxed density matrix as a starting point. The CCSD(T) total energies have also been considered a reference for all calculations.
All calculations have been performed using the \textit{Psi4}\cite{psi4} quantum chemistry package, except for full KS OEP calculations for which the \textit{ACESII}\cite{acesII} package was used. In the case of the latter, to solve the OEP equations, we have employed the finite-basis set procedure from \RRef{ivanov:1999:OEP}. For more computational details of the OEP procedure, we refer the reader to \RRef{grabowski:2014:jcp}. The WY calculations, in turn, were realized using the \textit{n2v}\cite{n2v} package combined with the \textit{Psi4} software.  For this method, we used a trust-exact algorithm implemented in SciPy \cite{virtanen2020scipy} for the optimization of the corresponding KS potential and tight convergence criteria set on the gradient norm (a convergence tolerance set to $10^{-6}$). 
As seed potential for the WY algorithm, we have used the Fermi-Amaldi potential~\cite{fermi:1934:fa} to ensure the correct $-1/r$ asymptotic behavior of the resulting XC potential. This feature is extremely important because the KS orbital energies (that enter the denominator in \Eq{eq:glpt2}) are very sensitive to the quality of the XC potential\cite{SmigaJCP2020, SOSa, Vignesh1}. However, a few initial tests performed for molecular systems revealed that all XC DFA results, including the second-order energy expression, are not sensitive to the choice of guiding potential in the WY procedure. This confirms the findings from \RRef{MDDFDE}. This should not be surprising since the choice of the guiding potential produces only a rigid shift of the eigenvalues spectrum that does not affect \Eq{eq:glpt2}. 

In all calculations, the uncontracted aug-cc-pVTZ \cite{unc-aug-cc-pvtz} atomic orbital basis set has been utilized to make a comparison on the same footing and to reduce basis set-related errors. The same basis set has been used for potential expansion to ensure a smooth, balanced solution for the WY and OEP methods. As suggested in a similar study\cite{MDDFDE}, the triple zeta quality basis set is sufficiently large to reach physically meaningful conclusions. 

Next, the orbitals and eigenvalue energies obtained from the WY calculations have been used to feed the second-order energy expressions, which have been implemented in an in-house code and interfaced with the \textit{Psi4} package using the Psi4NumPy\cite{psi4numpy} engine. For clarity of discussion, in the following, all WY results are labeled as @WY[HF], @WY[MP2], @WY[CCSD], and @WY[CCSD(T)] to underline the fact that these data give rise to a set of KS orbitals and eigenvalues that yield the same density as that obtained from a standard WFT: HF, MP2, CCSD and CCSD(T) calculation, respectively. The same notation, namely @HF, @PBE, and @PBE0, denotes the orbitals and eigenvalues obtained from HF, PBE, and PBE0 densities, respectively. Finally, the use of OEP SCF and GKS references has been denoted with @SCF and @GKS, respectively.

\subsection{B. Test cases}
To evaluate the performance and the errors of the different approaches, we applied them in several relevant contexts, focusing mainly on their application to real-world problems such as reaction energies and non-covalent interaction energies.
More specifically, we have considered:
 \begin{itemize}

\item \textbf{Total energies:} These have been calculated for the systems listed in Table I in \RRef{grabowski:2014:jcp} using the geometries indicated in that study. Although total energies are not very important in practical chemical applications, they are essential observables and are especially useful as indicators of the quality of DFA approximations.

\item  \textbf{Noncovalent interaction energies:}
we analyzed a few types of non-covalent molecular interactions like weak (Ar$_2$, He$_2$, NeHe, ArNe, Ne$_2$), dipole-dipole (H$_2$S-H$_2$S, H$_2$S-HCl, HCl-HCl), hydrogen-bonded (H$_2$O-H$_2$O, HF-HF, NH$_3$-NH$_3$). As for geometries, we have utilized those from \RRef{RefGeomTruhlar}. All quantities have been calculated without counterpoise corrections for basis set superposition error (BSSE). The ACMs data include the size-consistency correction from \RRef{vuckovic2018restoring}.

\item \textbf{Reaction energies:} these are quantities of primary interest in many chemical applications. To this end, we have selected nine closed-shell representative reaction energies (RE9) from \RRef{BozkayaSET} with the geometries from \RRef{g2_geom} (the list of reactions can be found e.g. in Tab. S16 in SI file). All quantities have been calculated without counterpoise corrections for basis set superposition errors.

 \item \textbf{Harmonium atoms and dissociation of H$_2$:} \rev{additionally, for the ACM class of functionals, we have tested their predictive power for the systems where strong correlation effects emerge, namely the Harmonium 
 atom \cite{PhysRev.128.2687} and the dissociation of H$_2$ using spin-restricted formalism. For the former, the calculations have been performed for $\omega \in [0.03 \div 1000]$ values. In this case, we have used an identical computational setup as in our previous study\cite{LUCISI,SCF-ISI}.}
 \end{itemize}

\subsection{C. Error statistics}

For each quantity and every data set, we have computed standard statistic error measurements, i.e., mean error (ME), mean absolute error (MAE), and mean absolute relative error (MARE). 
Furthermore, we have considered the following indicators:
\begin{equation}
  \text{RMAE}[\tilde{\mathit{R}}] =  \frac{\text{MAE}[\tilde{\mathit{R}}]}{\text{MAE}[\mathit{R}]}\; ,
  \label{eq:RMAE}
\end{equation}
which is the ratio of the MAE for a given method computed using the reference $\tilde{\mathit{R}}$ with the MAE of the same method but computed using the NeX reference. Note that this latter quantity is just the mean absolute error of the FD contributions.
The RMAE indicator allows for understanding the average error compensation effect between the FD and RD errors. In fact, when the RMAE $<$ 1, the method benefits from mutual RD and FD error compensation effects (the MAE is smaller when the approximate reference is employed); on the contrary, when the RMAE $>$ 1, the method RD and FD error contributions sum up (the MAE is bigger when the approximate reference is employed).

\section{III. Results and Discussions}
\label{sec:res}

In this section, we present the results obtained from the analysis of second-order correlation functionals, following the methodology outlined in Section II. Our focus is on describing and understanding the accuracy and limitations of these functionals in predicting key properties, including total energies, binding energies, and reaction energies.

\subsection{A. Total energies}
Table \ref{tab_totene} reports the MAE for total energies of the above-mentioned functionals using different references.
\begin{table*}[th!]
    \caption{\label{tab_totene}Mean absolute errors (in mE$_h$) for the total energies of various functionals with different reference sets of orbitals and eigenvalues. }
    \centering
    \resizebox{1\textwidth}{!}{
    \begin{tabular}{lrrrrrrrrr}
    \hline\hline
    Functional & @HF & @PBE0 & @PBE & @SCF & @WY[HF] & @WY[MP2] & @WY[CCSD] & @WY[CCSD(T)] \\
    \hline
    \multicolumn{9}{c}{\emph{ab initio} functionals}\\
GL2 & 19.92 & 68.90 & 118.98 & 124.14 & 98.54 & 113.43 & 110.10 & 111.68 \\
SC  & 19.92 & 14.37 & 12.31 & 15.04 & 18.46 & 13.82 & 14.74 & 14.33 \\
    \multicolumn{9}{c}{ACM functionals}\\
ISI & 46.09 & 25.04 & 57.94 & 55.10 & 46.00 & 55.22 & 53.37 & 54.28 \\
SPL & 44.47 & 31.60 & 68.69 & 65.92 & 54.40 & 65.27 & 62.95 & 64.06 \\
    \multicolumn{9}{c}{Double-hybrid functionals}\\
B2PLYP & 108.67 & 134.08 & 144.82 & 146.53 & 139.12 & 144.42 & 143.92 & 144.21 \\
PBE-QIDH & 66.59 & 84.15 & 95.06 & 97.68 & 92.94 & 96.23 & 96.03 & 96.16 \\
XYG3 & 116.15 & 141.88 & 153.11 & 154.62 & 151.61 & 154.74 & 154.63 & 154.73 \\
BL1P & 59.74 & 129.87 & 158.78 & 148.86 & 128.96 & 136.76 & 135.61 & 136.21 \\
    \hline\hline
    \end{tabular}}
\end{table*}
We can see that in all cases, the errors are quite large. This traces back to the fact that the second-order correlation expression has a large intrinsic FD error (@WY[CCSD(T)] column in the table), which amounts to more than 0.1 E$_h$ for the bare GL2 functional.  \rev{The individual $\Delta E_{FD}$ errors are reported in Fig. S2 in SI file. One can note that for GL2 functional the largest $\Delta E_{FD}$ are found for
the N$_2$ molecule (-0.25 E$_h$) whereas the smallest one for the He atom (-0.005 E$_h$).} A  slight improvement is obtained with the ACM functionals, which partially renormalize the GL2 behavior\cite{map_paper}. \rev{We note that ACM $W_\alpha$ curvature helps to reduce the GL2 overestimation in the cases where this is required (e.g., see ISI and SPL $W_\alpha$ N$_2$ integrand in Fig. S3 in SI), whereas it correctly preserves an almost linear behavior when GL2 is already good enough (e.g., see ISI and SPL $W_\alpha$ Ne integrand in Fig. S4 in SI).} In contrast, the DH functional formulation accumulates more errors in the FD alone, worsening the situation. \rev{In this case, the larger $\Delta E_{FD}$ errors can be noted for more systems (e.g., Mg, Ar, CO, Cl$_2$, N$_2$) despite the utilization of different DH DFA.} An exception is the SC functional, which takes advantage of the modified reference Hamiltonian and SC transformation of orbitals, implementing 
a much more effective perturbative correction to the total energy. 

The fact that the employed reference is crucial for the effectiveness of the second-order perturbation is confirmed by the observation that the GL2 expression indeed performs much better when it is used with @HF orbitals (i.e., MP2 is considered) since this is the basic framework where the second-order correction has been developed. Similar behavior is also observed for all the other functionals considered here, which inherit the behavior of the GL2 expression \cite{daas23}. It shall be further analyzed in terms of error compensation. On the other hand, using the KS reference, @PBE0 shows an exceptional performance for the ACMs, which is much better than that of the @HF orbitals. But @SCF, @PBE, and @WY( @WY[HF] being the best amongst the others) only have a small effect on functional performance.
This means that for all functionals, we can generally expect a relatively small contribution of the RD error, which cannot compensate for the FD error and, at times, simply worsens by adding onto it. @HF (as mentioned above) and @PBE0 are two exceptions since, in these cases, a larger RD error is found, and thus, a better compensation with the FD error is obtained. This is shown in Table \ref{tab_totene_errors}, where we report the mean ED and OD errors for the various cases.
\begin{table*}[th!]
    \caption{\label{tab_totene_errors} Mean OD and ED errors (in mE$_h$) for the total energies of various functionals with different reference sets of orbitals and eigenvalues.}
    \centering
    \begin{tabular}{lrrrrrrr}
    \hline\hline
    Functional & @HF & @PBE0 & @PBE & @WY[HF] & @WY[MP2] & @WY[CCSD] \\
    \hline
    \multicolumn{7}{c}{\textbf{Mean orbital-driven (OD) error}}\\
    \multicolumn{7}{c}{\emph{ab initio} functionals}\\
GL2 &  50.88 & 12.47 & -1.05 & 7.80 & -0.70 & 0.70 \\
SC  &  5.44 & 0.27 & -1.82 & 4.09 & -0.54 & 0.45 \\
    \multicolumn{7}{c}{ACM functionals}\\
ISI &35.34  &8.74 &0.69 &4.60 &-0.19 & 0.30\\
SPL & 39.50 &9.81 &0.24 & 5.52 &-0.37 &0.42\\
    \multicolumn{7}{c}{Double-hybrid functionals}\\
B2PLYP &  14.92&2.22 &1.00 &3.68 &0.08 &0.05\\
PBE-QIDH &14.43  & 3.49&3.50 &1.60 &0.31 &-0.16\\
XYG3 & 13.00 &3.24 &3.54 & 1.40& 0.32&-0.18\\
BL1P & 30.21 & -13.45& -18.18 & 3.66 &0.14 &0.00\\
&  & & & & &  \\
    \multicolumn{7}{c}{\textbf{Mean eigenvalue-driven (ED) error}}\\
    \multicolumn{7}{c}{\emph{ab initio} functionals}\\
GL2 &  80.73 & 30.32 & -6.24 &  5.34 & -1.05 & 0.89 \\
SC  &  0.15 & -0.23 & -0.20 &  0.05 & 0.02 & -0.04 \\
    \multicolumn{7}{c}{ACM functionals}\\
ISI &64.21  &22.40 &-4.41 & 3.78&-0.72 & 0.61\\
SPL & 67.93 &24.53 &-4.94 & 4.23 &-0.82 &0.69\\
    \multicolumn{7}{c}{Double-hybrid functionals}\\
B2PLYP &21.81  &8.08 & -1.62& 1.44&-0.28 &0.24 \\
PBE-QIDH &26.92  & 9.98&-2.00 & 1.78&-0.34 &0.29\\
XYG3 & 25.93 &9.61 & -1.92 & 1.71&-0.33 &0.28\\
BL1P & 54.30 & 19.78&-4.40 & 3.59& -0.69&0.60\\
    \hline\hline
    \end{tabular}
\end{table*}
{The SC method holds a very different behavior compared to the GL2-based functionals. We noticed exceptional behavior for @HF for the other functionals, but in contrast, there is error accumulation in the SC method. All the reference orbitals have worse performance than NeX.
Conversely, it appears that @WY[MP2] and @PBE have exceptional performance for the SC method, which had poor performance for GL2-based functionals.~Interestingly, it points to the fact that the FD error is much reduced for SC, and the MAE is very sensitive to the choice of reference orbitals. This behavior probably traces back to using the SC transformation, which mostly reduces the impact of the eigenvalues in the second-order energy formulation. This could also be confirmed from Table\ref{tab_totene_errors}; here, the OD dominance is strongly visible from the choice of orbitals. However, for the GL2-based functionals, the RD error receives comparable contributions from the ED and the OD terms.} This is an interesting finding since it confirms that density alone is not a sufficient descriptor for these orbital-dependent functionals. In particular, @HF and @WY[HF] calculations share the same density (what is shown in Tab. S1 in the SI file where we compare integrated density differences\cite{grabowski:2014:jcp}) yielding at the same time quite different RD (see  Fig. S1 in the SI file). This is a consequence of a large variation in ED error, which should be expected since the eigenvalues are known to be quite different but also have very different OD errors. This traces back to the fact that even though they sum up to yield the same density, the individual orbitals are different in the two cases. \\

\begin{figure}
\centering
\includegraphics[width=0.95\columnwidth]{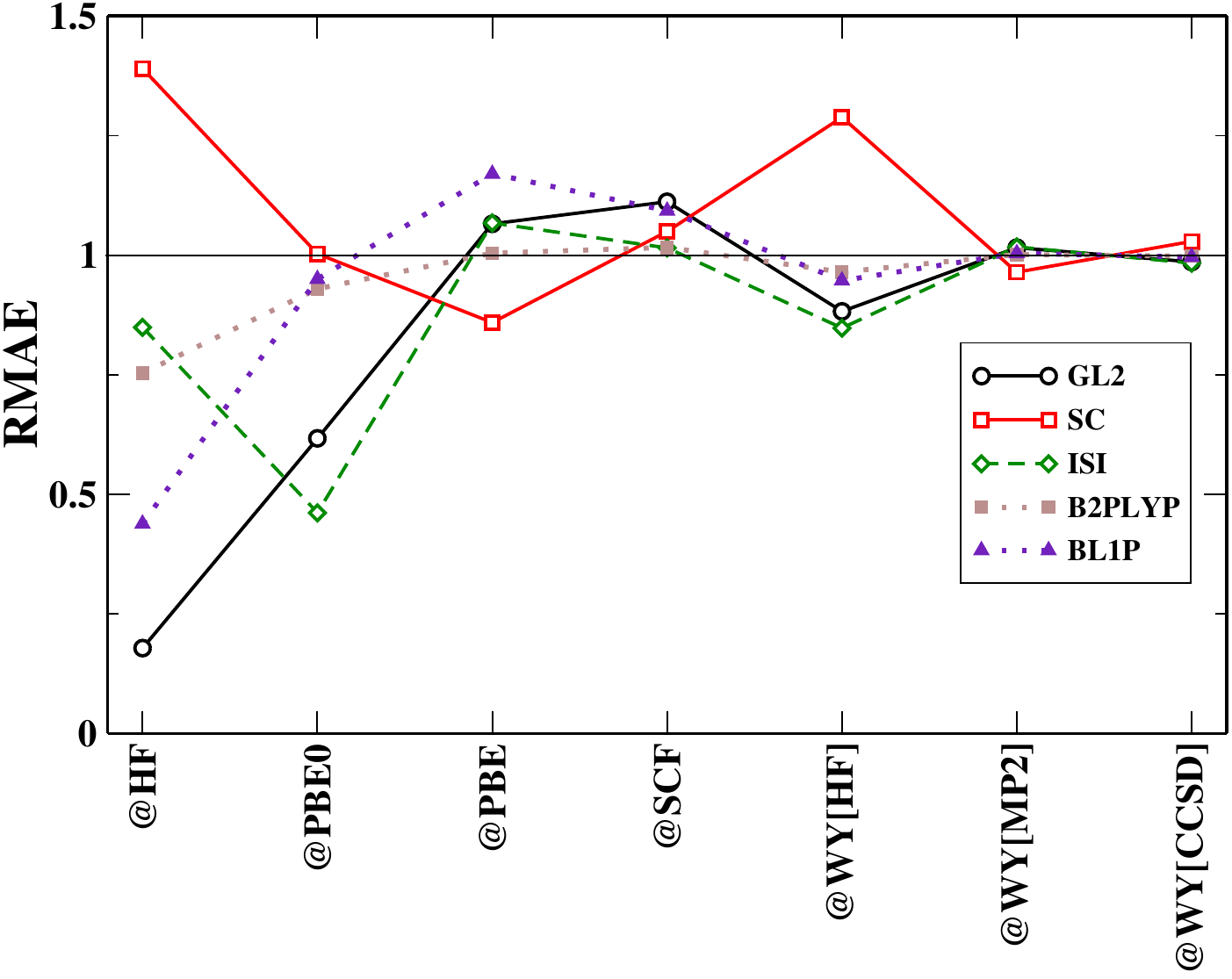}
\caption{\label{fig_totene_rmae}Relative mean absolute error (RMAE), Eq. (\ref{eq:RMAE}), for the total energies computed with different functionals and references. For clarity of the figure, we do not report SPL results, which are very close to ISI ones, and we do not report PBE-QIDH and XYG3 results, which are close to the other DH ones. Full results can be found in the supporting information (see Fig. S5). }
\end{figure}

Finally, in Fig. \ref{fig_totene_rmae}, we report the RMAEs relative to the total energy calculations performed with different functionals and references.
As we can expect from the previous analysis, a strong error compensation is found in all cases, except for the SC functional, when @HF orbitals are employed. Partially, the above trend is also true in the case of hybrid @PBE0 orbitals. It is the best choice for ACMs. On the other hand, using a KS reference is not very important, and only small variations in the RMAE can be observed. It is fascinating to note that the BL1P functional stays closer to the NeX solid line for any choice of reference (except @HF). This implies that RD's influence in this case is negligible, indicating that it has been formulated as having a smaller impact from the second-order term. Fig \ref{gks} depicts a special behavior for the GKS scheme in the DH; it points towards the benefit of error cancellation for all the DH cases.

\subsection{B. Binding energies}

In Table \ref{tab_bindene}, we report the MAEs for non-covalent binding energies as obtained from different methods and references.
\begin{table*}[ht]
 \caption{\label{tab_bindene} Mean absolute errors (in kcal/mol) for the binding energies of various functionals with different reference sets of orbitals and eigenvalues. }
    \centering
    \resizebox{1\textwidth}{!}{
    \begin{tabular}{lrrrrrrrrr}
    \hline\hline
    Functional & @HF & @PBE0 & @PBE & @SCF & @WY[HF] & @WY[MP2] & @WY[CCSD] & @WY[CCSD(T)] \\
    \hline
    \multicolumn{9}{c}{\emph{ab initio} functionals}\\
GL2 &  0.12 & 0.68 & 1.16 & 0.93 & 0.72 & 0.88 & 0.82 & 0.88 \\
SC  &  0.12 & 0.22 & 0.27 & 0.21 & 0.18 & 0.23 & 0.22 & 0.23 \\
    \multicolumn{9}{c}{ACM functionals}\\
ISI &  0.09 & 0.42 & 0.75 & 0.57 & 0.47 & 0.58 & 0.54 & 0.57 \\
SPL &  0.10 & 0.46 & 0.85 & 0.64 & 0.53 & 0.65 & 0.60 & 0.64 \\
    \multicolumn{9}{c}{Double-hybrid functionals}\\
B2PLYP & 0.40 & 0.21 & 0.25 & 0.23 & 0.31 & 0.26 & 0.26 & 0.25 \\
PBE-QIDH & 0.12 & 0.06 & 0.09 & 0.09 & 0.12 & 0.10 & 0.09 & 0.09 \\
XYG3 & 0.20 & 0.11 & 0.12 & 0.09 & 0.12 & 0.10 & 0.10 & 0.10 \\
BL1P &  0.09 & 0.20 & 0.29 & 0.31 & 0.25 & 0.29 & 0.26 & 0.27 \\
    \hline\hline
    \end{tabular}}
   
\end{table*}
Non-covalent binding energies are quite small energy differences and are, therefore, good quantities to investigate the fine effects of the choice of the reference in the functional performance. The table shows that the general trend for \emph{ab initio} and ACM functionals is similar to that already observed for the total energies. The FD error is much smaller in absolute terms \rev{(last column of Table \ref{tab_bindene})} than the total energies. However, it might be significant due to the small values of these energies. \rev{This can be seen in detail in Fig. S7 in the SI file, where we report the individual $\Delta E_{FD}$ error values for all functionals. } 
Here, we observe that ACM and DH have smaller FD to GL2, which was not the case for DH in total energy. The ACM and DH try to reduce the GL2 overestimation.
\rev{This is well seen in the cases where GL2 DFA exhibits the most significant FD errors, e.g., H$_2$S-HCl ($\Delta E_{FD} \approx$ 2.7 kcal/mol). The smallest FD errors are yielded by XYG3 (MAE = 0.10 kcal/mol) and PBE-QIDH DFAs (MAE = 0.09 kcal/mol).} It is important to note that the DHs (exception BL1P) trend has been flipped compared to the total energy, especially with the @HF reference, where we see a large MAE pointing towards error accumulation. The KS 
 reference @PBE0 orbitals have smaller MAE than @HF orbitals for DHs. Other choices of KS reference for DHs mainly do not impact or worsen the behavior. The \textit{ab initio} and ACM have the RD error comparable in magnitude to the FD error for binding energies for @HF reference. Table \ref{tab_bindne_errors} reports the mean ED and OD errors for the different functional and references.
{It is noteworthy that the SC shows a similar trend to GL2 and ACM, pointing out that there is a certain balance between the system and subsystems, which cancels the semi-canonical transformed effect. The OD still dominates in SC but is now less pronounced. The OD and ED contributions are on the same footing for all the GL2-based functionals.}

\begin{table*}[th!]
    \caption{\label{tab_bindne_errors} Mean OD and ED errors (in kcal/mol) for the binding energies of various functionals with different reference sets of orbitals and eigenvalues.}
    \centering
    \begin{tabular}{lrrrrrrr}
    \hline\hline
    Functional & @HF & @PBE0 & @PBE & @WY[HF] & @WY[MP2] & @WY[CCSD] \\
    \hline
    \multicolumn{7}{c}{\textbf{Mean orbital-driven (OD) error}}\\
    \multicolumn{7}{c}{\emph{ab initio} functionals}\\
GL2 &   0.35 & 2.50 & 3.53 & -0.13 & 0.09 & 0.12 \\
SC  &   -0.15 & -0.02 & 0.02 & -0.07 & 0.00 & -0.02 \\
    \multicolumn{7}{c}{ACM functionals}\\
ISI &0.49&2.25&3.15&-0.05&0.11&0.10 \\
SPL & 0.44&2.35&3.35&-0.07&0.10&0.11\\
    \multicolumn{7}{c}{Double-hybrid functionals}\\
B2PLYP & 0.16&0.78&0.94&-0.05&0.02&0.04 \\
PBE-QIDH &0.28&0.95&1.14&-0.02&0.04&0.05\\
XYG3 & 0.33&0.93&1.07&-0.01&0.03&0.06\\
BL1P &0.52&-1.04&-0.78&-0.03&0.06&0.10 \\
&  & & & & &  \\
    \multicolumn{7}{c}{\textbf{Mean eigenvalue-driven (ED) error}}\\
    \multicolumn{7}{c}{\emph{ab initio} functionals}\\
GL2 &  -1.14 & -2.70 & -3.25 & -0.08 & -0.10 & 0.18 \\
SC  &   0.01 & 0.01 & 0.02 & 0.02 & 0.00 & 0.01 \\
    \multicolumn{7}{c}{ACM functionals}\\
ISI &-1.05&-2.40&-2.97&-0.10&-0.11&-0.15\\
SPL & -1.06&-2.52&-3.13&-0.09&-0.11&-0.16\\
    \multicolumn{7}{c}{Double-hybrid functionals}\\
B2PLYP &-0.31&-0.75&-0.94&-0.02&-0.03&-0.05 \\
PBE-QIDH &-0.38&-0.92&-1.16&-0.03&-0.04&-0.06\\
XYG3 &-0.37&-0.89&-1.12&-0.02&-0.03&-0.06 \\
BL1P &-0.77&1.05&0.81&-0.05&-0.07&-0.12 \\
    \hline\hline
    \end{tabular}
\end{table*}

\begin{figure}
    \centering
    \includegraphics[width=0.95\columnwidth]{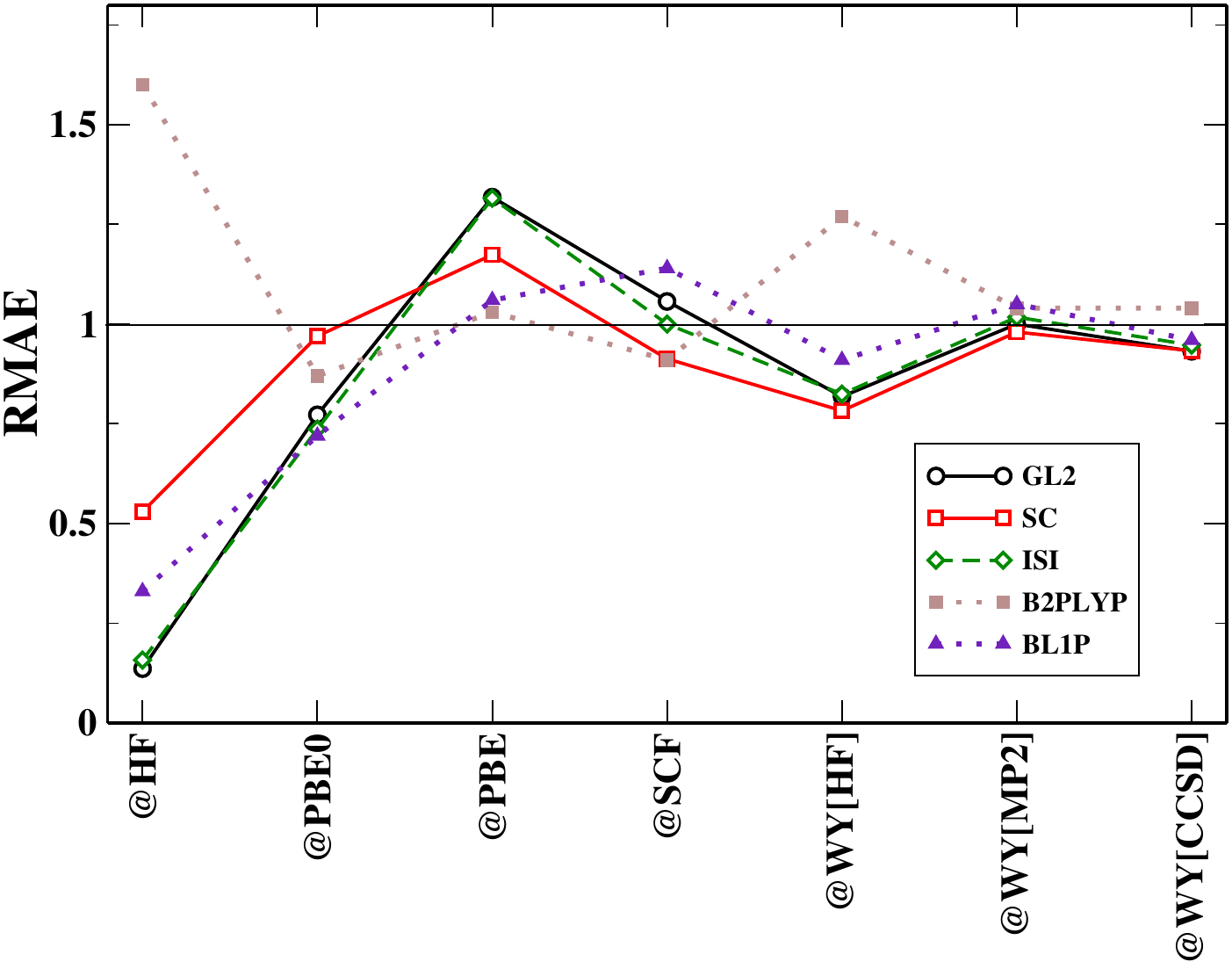}
    \caption{\label{fig_bindene_rmae} Relative mean absolute error (RMAE), Eq. (\ref{eq:RMAE}), for the binding energies computed with different functionals and references. For the clarity of the figure, we do not report SPL results here, which are very close to ISI ones, and we do not report PBE-QIDH and XYG3 results, which are close to the other DH ones. Full results can be found in the supporting information (see Fig. S8). }
\end{figure}

In Fig. \ref{fig_bindene_rmae}, we report the RMAEs for the binding energies obtained from various functionals and references. 
 Mostly, the choice of the reference has little effect on all the functionals, except when @HF orbitals and eigenvalues are considered. In this case, we have a general reduction of the MAE (RMAE < 1). However, the B2PLYP functional performs poorly with (RMAE >1); Table \ref{tab_bindne_errors} provides the logical explanation of how the OD and ED cancel each other, reducing the impact of RD and, thus, adding together with FD to have huge error accumulation. {BL1P functional follows a similar trend as total energy, staying closer to the NeX solid line for any choice of reference (except @HF). @PBE0 orbital carries the same trend as total energy, but the error cancellation is less dominant here. Interestingly, the GL2 and ACM follow 
 exactly similar trends for all references. \rev{Similar situation can be noted for FD errors reported in Fig. S7 in the SI file. This confirms the huge impact of the GL2 term in the ACM formulation.} Figure \ref{gks} for the GKS approach reports that all DH functionals (except BL1P) remain close to the NeX solid line. The scheme is beneficial with regard to error cancellation for the BL1P functional; the other DHs do not benefit much from it.}

\subsection{C. Reaction energies}
\begin{figure}
    \centering
    \includegraphics[width=0.95\columnwidth]{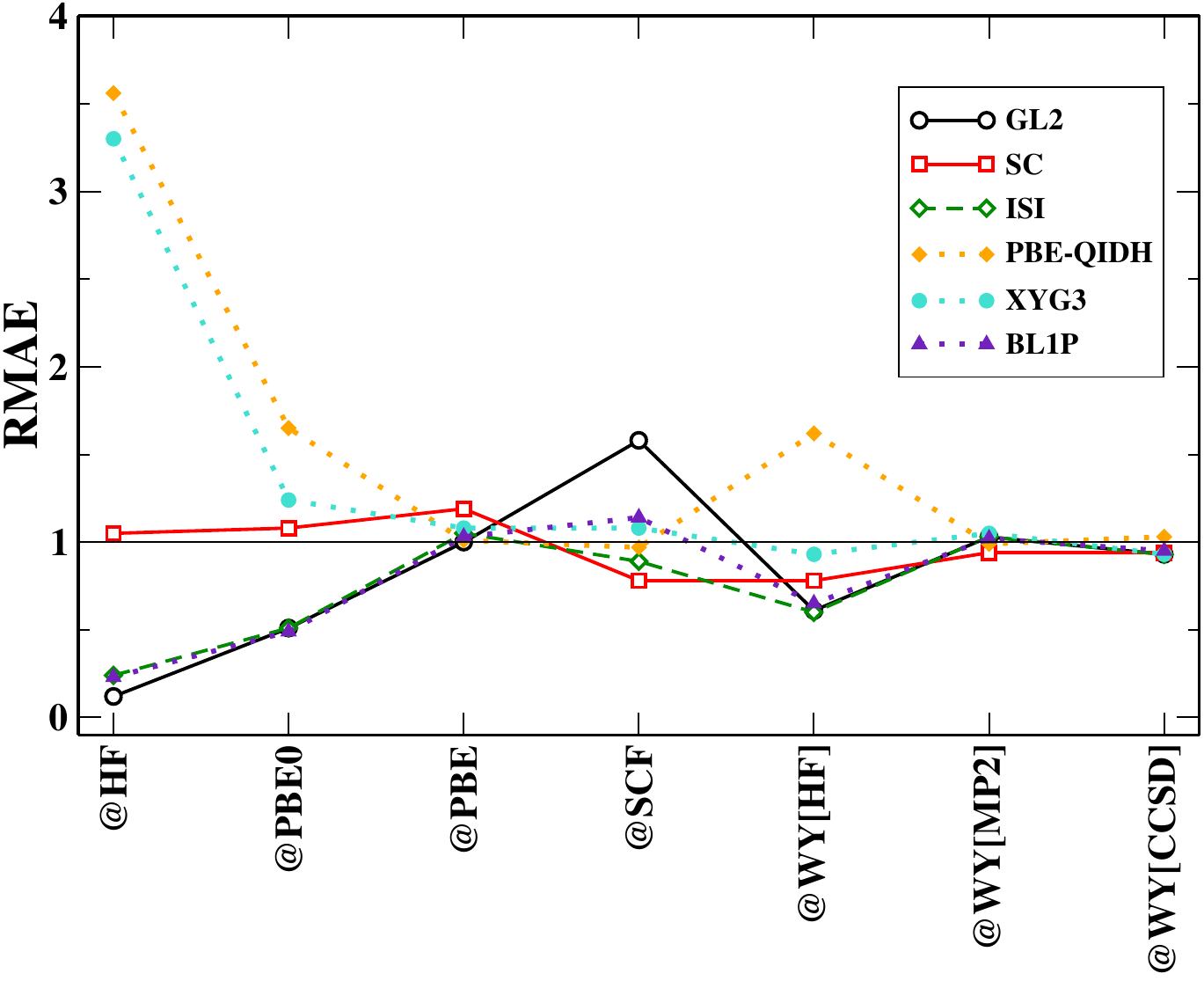}
    \caption{\label{fig_recene_rmae}Relative mean absolute error (RMAE), Eq. (\ref{eq:RMAE}), for the reaction energies computed with different functionals and references. For the clarity of the figure, we do not report SPL results, which are very close to ISI ones, and we do not report B2PYLP results, which are close to the other DH ones. Full results can be found in the supporting information (see Fig. S11).}
\end{figure}

\begin{figure}
    \centering
    \includegraphics[width=0.95\columnwidth]{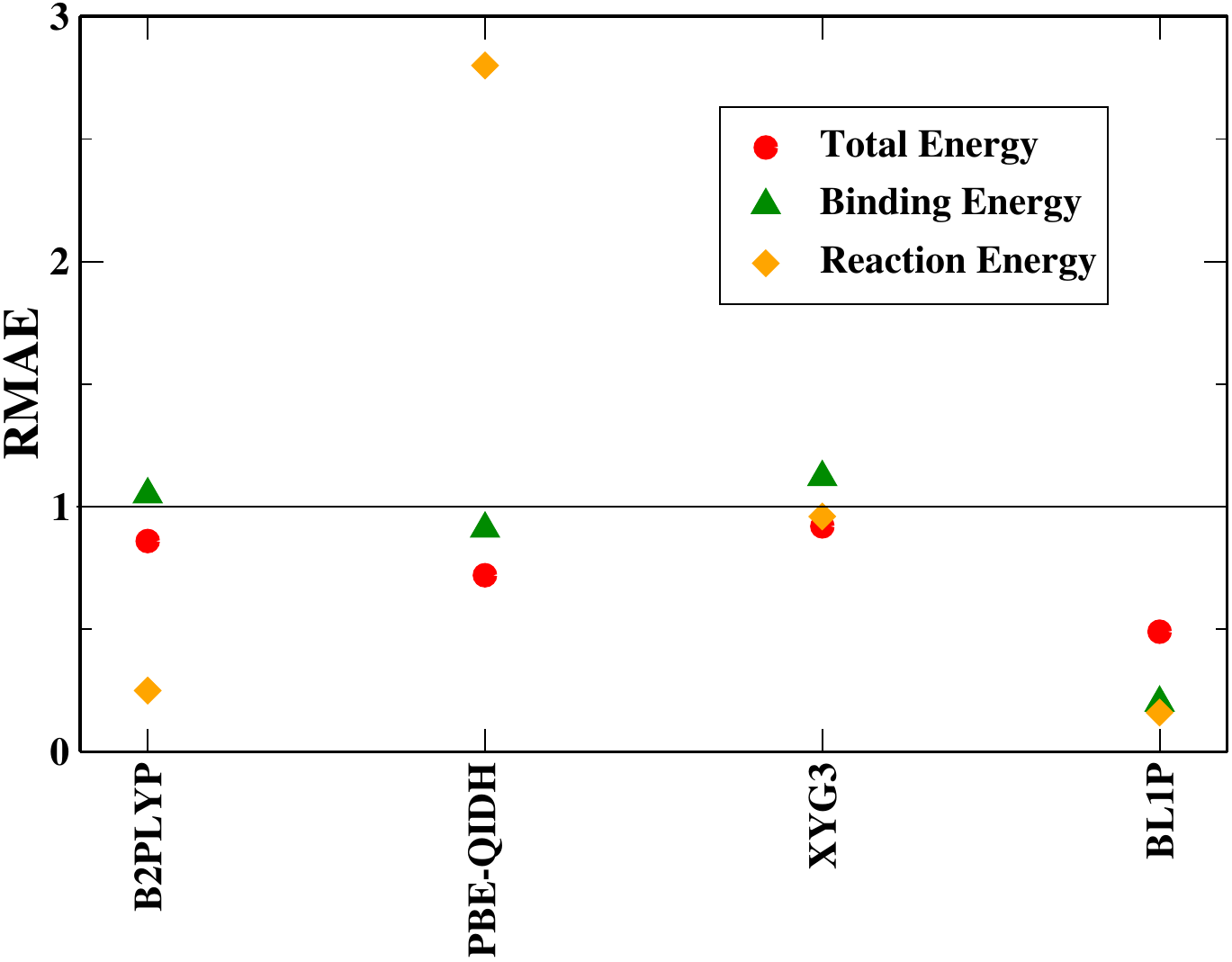}
    \caption{\label{gks}Relative mean absolute error (RMAE), Eq. (\ref{eq:RMAE}), for the GKS scheme computed for different DH cases, including total, binding, and reaction energies computed with different functionals and references. Full results can be found in the SI file. }
\end{figure}

{Let us now focus on Table \ref{tab_recene}, where we report the error statistics for several reaction energies computed for different choices of orbitals for the above-mentioned functionals. The GL2-based functionals have reduced FD errors compared to the total energy but not as small as binding energy. As mentioned earlier, the ACMs and DHs try to reduce the GL2 overestimation, which is also prominent here. \rev{This is shown in detail in Fig. S10 in the SI file, where we report the individual $\Delta E_{FD}$ error values for all functionals. One can note that again, XYG3 (MAE = 1.27 kcal/mol) and PBE-QIDH  (MAE = 1.43 kcal/mol) yield the smallest mean FD errors (last column of Table \ref{tab_recene}).} 
@HF, @PBE0, and @WY[HF] show smaller MAEs pointing to RD and FD error cancellation (exceptions are SC, PBE-QIDH and XYG3). This trend points to the fact that there is an interlink between the Hartree exchange contribution in the functional and the choice of orbitals. As mentioned for binding energy, the KS reference, @PBE0, is much better than @HF for DHs (exception BL1P). The SCF procedure can worsen the outcome, leading to error accumulation similar to the binding and total energy. The other choices of orbitals have smaller impacts.
Table \ref{tab_rec_errors} justifies how the ED and OD for these orbitals are of the same magnitude, canceling each other. Thus, we finally find smaller RDs. The OD dominance for SC still holds for reaction energies. An interesting trend was observed for ED error for @HF and @PBE0 orbitals; the values for ED for the two cases are very similar (except SC and BL1P). In \Fig{fig_recene_rmae}, the SC curve is always closer to the NeX in all cases. Rather, the other orbital choices have relatively little effect, as the graph illustrates. The curves are mostly seen approaching the NeX solid line, meaning the RD has a small magnitude. Similar is the BL1P behavior. PBE-QIDH and XYG3 both have error accumulation \rev{(RMAE > 1)} for @HF orbitals, which can be traced to the functional formulations of these DHs. The GKS scheme in \Fig{gks} shows that the PBE-QIDH functional \rev{faces the same problems (error accumulation with RMAE > 1)} for reaction energy. This also makes us realize that it is not well-optimized for these energies. The other DH mostly benefits from error cancellation. }

\rev{Lastly, we note that very similar RMAE trends for total, reaction, and binding energies have also been obtained in a larger basis set, namely uncontracted aug-cc-pVQZ\cite{unc-aug-cc-pvqz}. As an example, we report this data for GL2, ISI, and PBE-QIDH DFA in Figs S6, S9, and S13 in the SI file. This confirms that the triple zeta
quality basis set is sufficiently large to reach physically meaningful conclusions.}%
\begin{table*}[ht]
 \caption{\label{tab_recene} Mean absolute errors (in kcal/mol) for the reaction energies of various functionals with different reference sets of orbitals and eigenvalues. }
    \centering
    \resizebox{1\textwidth}{!}{
    \begin{tabular}{lrrrrrrrrr}
    \hline\hline
    Functional & @HF & @PBE0 & @PBE & @SCF & @WY[HF] & @WY[MP2] & @WY[CCSD] & @WY[CCSD(T)] \\
    \hline
    \multicolumn{9}{c}{\emph{ab initio} functionals}\\
GL2 & 2.19  & 9.27 & 18.19&  28.64& 11.02&18.71&16.88&18.13 \\
SC  & 2.19&2.26&2.50&1.63& 1.63&2.27&1.97&2.09 \\
    \multicolumn{9}{c}{ACM functionals}\\
ISI & 2.20&4.64&9.59&8.12&5.49&9.41&8.46&9.12  \\
SPL & 2.73&4.71&10.79&10.92&6.41&10.60&9.50&10.18 \\
    \multicolumn{9}{c}{Double-hybrid functionals}\\
B2PLYP &1.78&1.62&3.39&3.49&1.88&3.32&3.12&3.27  \\
PBE-QIDH & 5.10&2.36&1.44&1.39&2.32&1.42&1.47&1.43 \\
XYG3 & 4.19&1.58&1.37&1.37&1.17&1.33&1.18&1.27\\
BL1P & 2.20&4.72&9.99&11.07&6.32&9.89&9.20&9.67  \\
    \hline\hline
    \end{tabular}}
   
\end{table*}

\begin{table*}[th!]
    \caption{\label{tab_rec_errors} Mean OD and ED errors (in kcal/mol) for the reaction energies of various functionals with different reference sets of orbitals and eigenvalues.
   }
    \centering
    \begin{tabular}{lrrrrrrr}
    \hline\hline
    Functional & @HF & @PBE0 & @PBE & @WY[HF] & @WY[MP2] & @WY[CCSD] \\
    \hline
    \multicolumn{7}{c}{\textbf{Mean orbital-driven (OD) error}}\\
    \multicolumn{7}{c}{\emph{ab initio} functionals}\\
GL2 &  -12.04&-3.88&-2.42&-2.42&-0.06&-0.76  \\
SC  & -1.44&-0.75&-0.61&-0.80&0.16&-0.07  \\
    \multicolumn{7}{c}{ACM functionals}\\
ISI & -6.69&-2.04&-1.57&-1.12&-0.27&-0.33 \\
SPL &-8.16&-2.58&-1.85&-1.56&-0.18&-0.44 \\
    \multicolumn{7}{c}{Double-hybrid functionals}\\
B2PLYP & -2.96&-0.66&-0.14&-0.41&-0.18&-0.07 \\
PBE-QIDH &-3.51&-0.75&-0.11&-0.42&-0.24&-0.07\\
XYG3 &-3.24&-0.55&0.10&-0.30&-0.23&-0.07 \\
BL1P & -7.43&-2.12&-1.13&-1.15&-0.33&-0.29\\
&  & & & & &  \\
    \multicolumn{7}{c}{\textbf{Mean eigenvalue-driven (ED) error}}\\
    \multicolumn{7}{c}{\emph{ab initio} functionals}\\
GL2 &-5.09&-5.06&0.93&-2.67&0.78&0.12  \\
SC  &  -0.29&-0.09&0.02&-0.28&-0.01&-0.04 \\
    \multicolumn{7}{c}{ACM functionals}\\
ISI & -3.21&-2.90&0.32&-1.55&0.59&0.04\\
SPL & -3.60&-3.54&0.38&-1.88&0.66&0.03\\
    \multicolumn{7}{c}{Double-hybrid functionals}\\
B2PLYP & -1.37&-1.36&0.22&-0.69&0.24&-0.01\\
PBE-QIDH &-1.70&-1.68&0.27&-0.85&0.30&-0.01\\
XYG3 &-1.64&-1.62&0.26&-0.82&0.29&-0.01 \\
BL1P & -3.42&-3.40&0.81&-1.71&0.60&-0.01\\
    \hline\hline
    \end{tabular}
\end{table*}

\subsection{D. Harmonium atom and H$_2$ dissociation}
\rev{
In this section, we investigate the predictive power of ACM formulas for the systems where strong correlation effects emerge using two simple toy examples, i.e., H$_2$ dissociation (with spin-restricted formalism) and harmonium atom model, where FCI densities are relatively simply available. We note for these two systems, the HOMO-LUMO gap closes (when $\omega \rightarrow 0$ and $R/R_0 \rightarrow \infty$ for harmonium atom and H$_2$, respectively), causing the diverging of GL2 term. Consequently, all GL2-based DH and \textit{ab initio} DFT approximations fail in this regime. However, this is not the case for ACM approximations where the GL2 term is regularized using a non-linear $W_\alpha$ integrated formula in \Eq{e15}. For the sake of clarity, we only consider quantities obtained from @WY[HF] and @WY[CCSD] calculations, which are mostly suited for this construction\cite{gISI2}.}

\begin{figure}[h!]
    \centering
    \includegraphics[width=\columnwidth]{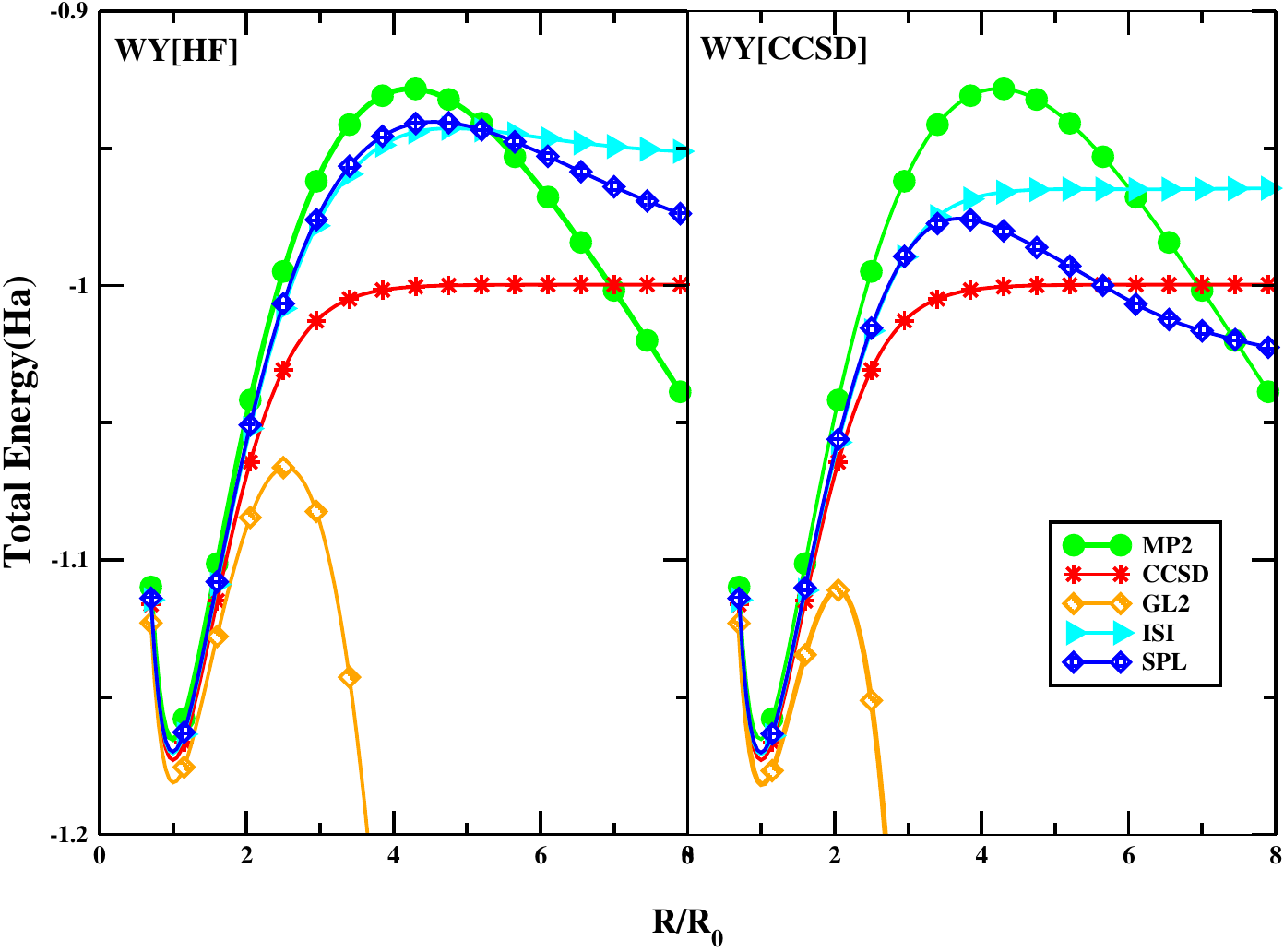}
    \caption{The total energy of the H$_2$ stretching computed using @WY[CCSD] and @WY[HF] orbitals for ISI and SPL functionals using the hPC model for the strong-interaction functionals. We report MP2 and CCSD/FCI data obtained in the same basis set for comparison.}
    \label{fig_h2}
\end{figure}

First, in \Fig{fig_h2}, we analyze the potential energy surface for the H$_2$ dissociation with spin-restricted formalism using KS orbitals generated from HF and CCSD/FCI densities \rev{for every single interatomic separation $R/R_0$.} We also report CCSD/FCI, MP2, and GL2 data for comparison. 
\Fig{fig_h2} highlights a few important features: i) in the right panel, the ISI and SPL, as well as GL2 results, do not carry (by construction) any RD error; ii) in the left panel, in turn, the performance is a consequence of mutual RD and FD error cancellation; iii) at equilibrium distance, all methods are quite accurate. The highest accuracy is obtained from the ISI formula;  iv) in the region around 4-8 a.u., a large repulsive bump emerges in the case of SPL and ISI, which is related to deficiencies of XC energy expression to describe fully the regions where static and dynamic correlation effects interplay\cite{fuchs2005desc,zhang16,vuckovic2016XC,vuckovic2017simple};  v) at the dissociation limit (where  $E_c^{GL2}=-\infty$), both methods asymptotically tend to a constant, in contrast to GL2, which diverges. This was discussed in detail in \RRef{gISI2}. With @WY[HF] input quantities, the SPL gives an almost perfect agreement with exact data in this limit\cite{gISI2}. Using NeX quantities, the SPL loses its accuracy, yielding lower energy, whereas ISI shows improved behavior. This indicates that, in this regime, the behaviors of ACMs strongly depend upon the quality of the approximations for $W_\infty$ and $W'_\infty$. We note, however, that once the exact SCE $W_\infty$ and $W'_\infty$ ingredients are used, both expressions shall yield the exact result\cite{gISI2}.

\begin{figure}[h!]
    \centering
    \includegraphics[width=\columnwidth]{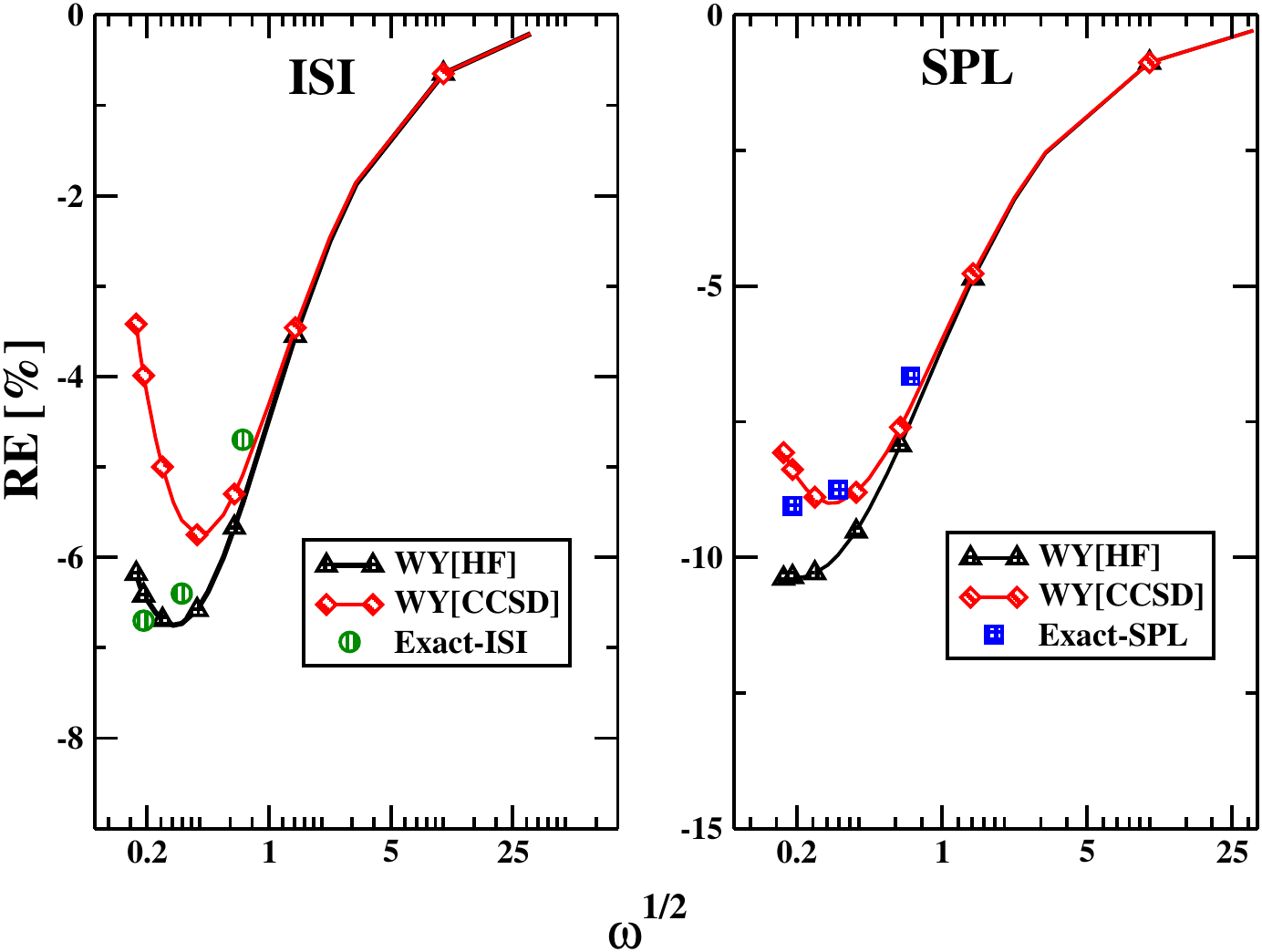}
    
        \caption{Relative error on correlation energies of harmonium atoms for various values of $\omega$ computed at @WY[CCSD] and @WY[HF] orbitals for ISI and SPL functionals using the hPC model for the strong-interaction functionals. The exact ISI and SPL values are taken from Ref. \cite{kooi2018local}, and are obtained by inserting exact densities into the ISI and SPL functionals, including the exact treatment (SCE) of the strong-interaction limit.}
    \label{fig_hook}
\end{figure}

In \Fig{fig_hook}, we show the relative errors (in \%) on 
correlation energies given by SPL and ISI functionals for Harmonium atom within a broad interval of frequencies $0.03\le \omega \le 1000$. As mentioned, the calculations used @WY[HF] and NeX @WY[CCSD] input quantities.

For the tighter bound electrons ($\omega\ge 1$) until the high-density limit ($\omega\gtrapprox 100$), one can note the lack of dependence upon the choice of reference orbitals and eigenvalues for both energy expressions. In the strongly correlated regime (i.e., $\omega\le 0.5$), the relevance of orbital choice is much more important. As one can note, both energy expressions overestimate the FCI data in this region, although the effect is less pronounced in the case of ISI DFAs. This is probably due to the inclusion of both $W_\infty$ and $W'_\infty$ terms in the DFA formula. Moreover, in the case of ISI ACM, the better agreement with exact\cite{kooi2018local} data is achieved for @WY[HF], whereas for SPL energy expression for NeX orbitals. Similar observations have also been made regarding the full self-consisted realization of ACMs\cite{SCF-ISI}. To investigate this in more detail in \Tab{tab_W} we report the error on ACM input ingredients $\mathbf{W}=(W_0,W'_0,W_\infty,W'_\infty)$ computed using @WY[HF] and @WY[CCSD] orbtitals. We can note that in the case of the latter (no RD error), the main source of error comes from  $W'_0$  and $W'_\infty$ terms, which is especially visible in the strong interaction limit. The error on  $W'_0$  is probably related to the insufficiency of the basis set to describe this term correctly. We note that reference data have been calculated at CBS limit\cite{kooi2018local}. The large error on $W'_\infty$, in turn, is probably related to the hPC formula, which was parametrized solely with respect to $\omega = 0.5$ exact $W_\infty$ and $W'_\infty$ values.  

In the case of @WY[HF], due to the inclusion of RD error, one can observe a significant increase of average relative errors for $W_0$ as well as  $W'_0$ and $W_\infty$ whereas for $W'_\infty$ we see the opposite trend. Nevertheless, in the case of ISI, one observes the bettering of predictions. It should be considered that this is just an error cancellation between the approximated electronic density and
the (approximated) hPC model. In the case of SPL (where we lack dependence on $W'_\infty$), we note the best agreement with exact data for @WY[CCSD]. This confirms that their behaviors strongly depend on the quality of the approximations used for $W_\infty$ and $W'_\infty$.

\begin{table}[h!]
\caption{The absolute relative errors and mean absolute relative errors (in \%) on  input ingredients $\mathbf{W}=(W_0, W'_0, W_\infty, W'_\infty$ ) computed using @WY[HF] and @WY[CCSD] input quantities with respect to exact data from Ref. \cite{kooi2018local}. }
\begin{tabular}{ccccc}
\hline \hline
$\omega$ & $W_0$ & $W'_0$ & $W_\infty$ & $W'_\infty$ \\ \hline
 &  & @WY[CCSD] &  &  \\ \hline
0.0365373 & 0.01\% & 2.24\% & 0.15\% & 7.65\% \\ 
0.1 & 0.01\% & 2.19\% & 0.10\% & 3.67\% \\ 
0.5 & 0.01\% & 2.13\% & 0.04\% & 0.62\% \\ 
 &  &  &  &  \\ 
MARE & 0.01\% & 2.19\% & 0.10\% & 3.98\% \\ \hline
 &  & @WY[HF] &  &  \\ \hline
0.0365373 & 1.57\% & 5.13\% & 1.40\% & 6.50\% \\ 
0.1 & 0.55\% & 3.65\% & 0.44\% & 2.87\% \\ 
0.5 & 0.04\% & 2.52\% & 0.01\% & 0.25\% \\ 
 &  &  &  &  \\ \hline
MARE & 0.72\% & 3.77\% & 0.61\% & 3.20\% \\ \hline \hline
\end{tabular}
\label{tab_W}
\end{table}

\section{IV. Conclusions}

This study analyzed the limitations of second-order correlation functionals through a detailed analysis of functional, orbital, and eigenvalue-driven errors, focusing on their performance in various contexts, such as total, binding, and reaction energies. The results indicate that second-order functionals, including GL2, ACM, and DH, show considerable error sensitivity based on the choice of reference orbitals and eigenvalues. \rev{An important role is thus also played by the reference, which was originally used to develop the functionals, as this can have an important impact on its optimization. This is evident, for example, looking at the difference between BL1P, which was based on HF orbitals, and other DHs.} A notable observation is the importance of error cancellation between FD and RD errors and, more importantly, the further decomposition of RD errors into OD and ED terms, which is important in further understanding the main source errors in second-order DFAs. 

For most functionals, the performance for various trial orbitals is mainly related to the orbitals spectrum (i.e., the ED term), not strictly to the quality of input density. One notable exception is the SC method because, in this case, the HOMO-LUMO gaps obtained from the SC transformed $H_0$ are very similar despite the orbital used. Thus, the main source of errors is related chiefly to OD error. For the DH DFAs, the total energy performance follows the GL2 trend, however, for other test sets, the trends are not so obvious and depend on multiple factors, i.e., the choice of trial orbitals (including the size of the HOMO-LUMO gap), mutual error cancellation effects between semi-local and \textit{ab initio} parts of DFA. Nonetheless, for DH DFAs, one can see that mostly these functionals benefit from GKS realization.
In contrast, the full @SCF realization of second-order DFAs usually leads to an overestimation of error, which indicates that the FD and RD add up with each other. However, FD is still the dominant contributor that governs their performance\cite{SCF-ISI,DHOEPSmiga2016, DHRSOEP}. It is also important to note that amongst all the choices for orbitals (mostly), HF and PBE0 are the best choices for the functionals due to the error cancellation between RD and FD. \rev{We remind, however, that since we remained confined to KS-DFT, employing HF orbitals can be considered an extreme hybrid strategy. As already stated, using a different perspective (e.g., perturbation theory from the HF reference) would lead to a very different partitioning of the error contributions.}

For future directions, improving the performance of these functionals will likely involve refining the methods for calculating reference orbitals and eigenvalues, especially with respect to OD and ED errors, to optimize the error balance. As an alternative path, one could consider working within a well-defined reference to develop novel energy expressions in ACM and DH, minimizing the intrinsic FD error. \rev{In this context, the study of large or metallic systems where the GL2 term gives divergent behavior could provide additional important insight. Moreover, } exploring alternative partitioning of Hamiltonian into $H_0$ and perturbation (as demonstrated with the SC approach in Appendix A) may also hold the promise of better error management. This could be an interesting strategy, especially in combination with ACM methods, where an improvement of the initial slope of the AC curve will surely bring a significant improvement in the overall performance.  All these improvements would contribute significantly to the ongoing development of density functional approximations aimed at achieving chemical accuracy.

\section*{Supporting Information}

The tables with raw data and additional figures supporting the analysis. 

\section*{Data Availability}

The data that support the findings are published within this study.

\section*{Acknowledgements}

S.\'S. thanks to the Polish National Science Center for the partial financial support under Grant No. 2020/37/B/ST4/02713.

\section{Appendix A: Overview of second-order dependent \\ XC density functional methods}
\label{Sec1}

In this section, we briefly review different approaches based on the perturbative second-order correlation grounded within the KS-DFT framework.

\subsection{1. The \textit{ab initio} DFT functionals}
The simplest manner to include correlation effects into \textit{ab initio} class of functional is via the utilization of second-order correlation energy expression\cite{grabowski:2002:OEPP2,mori-sanchez:2005:oeppt2,engel:2005:oeppt2} which for non-canonical orbitals (this is the case for KS-DFT or any inverse method)\cite{lauderdale:1992:GMBPT,bartlett:1995:coupled,grabowski:2002:OEPP2,gorling:1995:IJQCS} takes the form of the GL2 energy expression\cite{gorling:1994:OEP,ivanov:2003:P2}
\begin{equation}\label{eq11}
E_c^{GL2} = E_S^{GL2} + E_D^{GL2}.
\end{equation}
where $E_S^{GL2}$ and $E_D^{GL2}$ are single- and double-excited terms, respectively,
\begin{eqnarray}\label{eq:glpt2}
\nonumber
E_S^{GL2} &=& \sum_{ai} \frac{\left\lvert f_{ia}
\right\rvert^2}{\varepsilon_{i}^{} - \varepsilon_{a}^{}},\\ \label{abuk:glpt2-d}
E_D^{GL2} &=& \frac{1}{4}\sum_{abij} \frac{\left \lvert \left \langle ij \| ab \right \rangle \right
\rvert^2}{\varepsilon_{i}^{} + \varepsilon_{j} - \varepsilon_{a} -
\varepsilon_{b}^{}} \; .
\end{eqnarray}
Here, the standard notation for the antisymmetrized two-electron integrals is used
\begin{equation}
 \abraket{ij}{ab} = \braket{ij}{ab} - \braket{ij}{ba} \;.
\end{equation}
The latter term has the same form as the M\o{}ller-Plesset (MP2) correlation energy
expression evaluated using KS quantities, namely KS eigenvalues ($\varepsilon_{p}$) and orbitals $\phi_p(\R)$. 
In the case of a single excited term, the $f_{pq}$ denotes the elements of the Fock matrix, defined in terms of KS quantities 
\begin{align}
f_{pq}= h_{pq} - \sum_j \abraket{pj}{qj} \; ,
\end{align}
where  $h_{pq}$ are the matrix elements of the core Hamiltonian describing its kinetic energy and potential energy in the field of the nuclei.
The \Eq{eq:glpt2} emerges from the natural decomposition of total Hamiltonian ($H = H_0 + V$) into perturbation~$V$ and zeroth-order part $H_0$ which for GL2 energy expression is given by the sum of KS one-particle Hamiltonians
\begin{align}
H_0^{\mathrm{GL}} =  \sum_p
\varepsilon_p^{\mathrm{KS}} \{ p^\dagger p \},
\label{eq:H0gl2}
\end{align}
where $\{p^\dagger p\}$ denotes the normal product of spinorbital second-quantized operators.  The GL2 correlation energy expression combined with exact-exchange (EXX) energy expression
\begin{equation}
E_\text{x}^{\text{EXX}} = - \frac{1}{2} \sum_{i,j}
\braket{ij}{ji},
\label{eq:exx1}
\end{equation}
defines the OEP-GL2 XC functional ($E_{xc} = E_\text{x}^{\text{EXX}} + E_c^{GL2}$). As was noted in many studies, this choice of XC functional leads to a large overestimation of correlation effects\cite{grabowski:2002:OEPP2,bartlett:2005:abinit2,engel:2005:oeppt2,schweigert:2006:pt2,mori-sanchez:2005:oeppt2}, namely correlation energy, correlation potentials, correlated
density, and leads to problems with convergence in many cases\cite{grabowski:2011:jcp,bartlett:2005:abinit2,mori-sanchez:2005:oeppt2,grabowski:2013:molphys,grabowski13,grabowski:2014:jcp}. We note that some of the convergence issues might be resolved by a proper solution of OEP equations\cite{OEP_AQC} or by rescaling \Eq{eq:glpt2}\cite{grabowski:2014:jcp, SCSIP,KoderaGL2}.

In \RRef{bartlett:2005:abinit2}, it was noted that the choice of $H_{0}$ (given by \Eq{eq:H0gl2}) is not optimal, because
it forces the perturbation correction to retain diagonal elements of $H_0$. Thus, in order to solve some deficiencies of GL2 partitioning the alternative choice for $H_0$ was introduced \cite{bartlett:1995:coupled}
\begin{align}
H^{\text{SC}}_{0} &=\sum_{p}f_{pp}\{p^{\dagger }p\}, \label{H0f}\\
V &=\sum_{ai} f_{ai}\{a^{\dagger }i+i^{\dagger }a\}+W,
\end{align}
where $W$ is the two-particle term
\begin{equation}
 W = \frac{1}{4} \sum_{pqrs} \langle pq \| rs \rangle \{p^\dagger q^\dagger r s\}.
\end{equation}
Here, the SC transformation was enforced to keep the expression 
invariant concerning any unitary transformation among the occupied and/or virtual orbitals~\cite{bartlett:2005:abinit2}, and additionally to remove the off-diagonal ($f_{ij}$ and $f_{ab}$) elements from  $H_0$. This leads to the alternative second-order correlation energy expression, which, combined with \Eq{eq:exx1} can be used for any arbitrary set of orbitals
\begin{eqnarray}\label{eq:SC}
\nonumber
E_S^{\text{SC}} &=& \sum_{ai} \frac{\left\lvert f_{ia} \right\rvert^2}{f_{ii} -
f_{aa}},\\ \label{abuk:mbpt2f-d}
E_D^{\text{SC}} &=& \frac{1}{4} \sum_{abij} \frac{\left \lvert \left \langle ij \| ab \right \rangle \right
\rvert^2}{f_{ii} + f_{jj} - f_{aa} - f_{bb}}.
\end{eqnarray}
One can note that GL2 and SC correlation energy expressions reduce to MP2 counterparts if
canonical Hartree-Fock (HF) orbitals are used to feed \Eq{eq:glpt2} and \Eq{eq:SC}. As was shown in many studies\cite{bartlett:2005:abinit2,grabowski:2011:jcp,lotrich:2005:vdw,bartlett:2005:abinit3,
grabowski:2007:ccpt2,grabowski:2008:ijqc,SmigaJCP2020} the SC choice of $H_0$ provides much more stable results in comparison to GL2 variant.
%


\subsection{2. The ACM functionals}
The adiabatic connection (AC) formalism \cite{ACKS,langreth75,gunnarsson76,ACKS,savin03} provides a rigorous path of construction for XC functionals. The AC XC functional definition is based on the coupling constant ($\alpha\ge 0$ also known as interaction strength) integral formula~\cite{langreth75,savin03}
\begin{equation}
E_{xc}[\rho]=\int_0^1~d\alpha~W^{\alpha}_{xc}[\rho]
 \label{eq:ac1}
\end{equation}
with integrand defined as $W^{\alpha}_{xc}[\rho]=\langle\Psi^\alpha[\rho]|\hat{V}_{ee}|\Psi^\alpha[\rho]\rangle-U[\rho]$, where $\hat{V}_{ee}$ is the Coulomb operator, $U[\rho]$ is the Hartree energy, and $\Psi^\alpha[\rho]$ is the $\alpha$-dependent electronic
wave function which yields for any value of $\alpha$ the density $\rho(\R)$. The \Eq{eq:ac1} connects a non-interacting KS single particle system ($\alpha=0$) with a real, fully interacting one ($\alpha=1$). This provides the exact and formal definition of the XC functional. Unfortunately, the analytical formula for $W^{\alpha}_{xc}[\rho]$ is unknown; thus, usually, one tries to model a working expression by considering the known limits i.e. weak
\begin{eqnarray}\label{weaklim}
W^{\alpha\rightarrow 0}_{xc} & \sim & W_0 + \alpha W'_0 + \cdots \; ,
\end{eqnarray}
and strong
\begin{eqnarray}
W^{\alpha\rightarrow \infty}_{xc} & \sim & W_\infty + \frac{1}{\sqrt{\alpha}}W'_\infty + \cdots 
\end{eqnarray}
interaction limits. These have been directly employed to construct high-level XC functionals based on AC models (ACM)\cite{isi,isierr,seidl2000density,gorigiorgi09,SeiPerLev-PRA-99,liu09,ernzrehof99,teale10, gISI, gISI2}. 
The general form of ACMs reads as
\begin{equation}\label{e15}
E_{xc}^{ACM} = {\cal F}^{ACM}(\mathbf{W}) = \int_0^1 d\alpha W_\alpha^{ACM}(\mathbf{W})\;,
\end{equation}
where ${\cal F}^{ACM}$ is a non-linear function defining the energy expression (see e.g. Appendix A in \RRef{potISI})  and $\mathbf{W}=(W_0,W'_0,W_\infty,W'_\infty)$, with
$W_0=E_\text{x}^{\text{EXX}}$ being the exact exchange energy (see \Eq{eq:exx1}),
$W'_0=2E_c^{\rm GL2}$ being twice the GL2
correlation energy \cite{gl2} (see \Eq{eq:glpt2}), and $W_\infty$ and $W'_\infty$
being the indirect part of the minimum expectation value
of the electron-electron repulsion for a given density and the potential energy of coupled zero-point oscillations around this minimum, respectively \cite{seidl07,gorigiorgi09}. We note that $W_\infty$ and $W_\infty'$ are highly non-local density-dependent functionals, described exactly by the strictly-correlated electrons (SCE) limit \cite{seidl07,gorigiorgi09,MalMirCreReiGor-PRB-13}, and their exact evaluation in general cases is a non-trivial problem. Approximately, these
can be modeled by semilocal gradient expansions (GEA) derived within the point-charge-plus-continuum (PC) model~\cite{seidl2000simulation,seidl2000density,perdew2001exploring,seidl2010adiabatic}, the generalized gradient approximation (GGA) such as mPC\cite{Luc_mISI1}, hPC\cite{SCF-ISI} or meta-GGAs level of theory\cite{seidl2000density, ACSCTPSS}.

Considering the weak interaction limit expression, Eq. (\ref{weaklim}), the ACM approach can also be interpreted as a renormalization of the GL2 energy expression \cite{map_paper}. In fact, in the AC framework, the GL2 energy is given by the integration of a straight line, whereas the main action of the ACM model is to introduce a proper curvature into this line, such as to reduce the GL2 correlation overestimation and recover correct correlation energy.
In the past few years, several ACMs have been tested for various chemical applications, showing promising results \cite{fabianoisi16,isigold,LUCISI, gISI2}, especially in the description of non-covalent interactions \cite{spl2,vuckovic18}.
\rev{Here, we focus on two specific ACMs formulas, namely the interaction strength interpolation (ISI) \cite{isi} and Seidl-Perdew-Levy (SPL) \cite{SeiPerLev-PRA-99}. The ISI XC formula reads
\begin{equation}
E_{xc}^{ISI} = W_\infty +  \frac{2X}{Y}[\sqrt{1+Y} -1 -Zln(\frac{(\sqrt{1+Y}+Z)}{1+Z}] 
\end{equation} \\
with
\begin{equation}
X = \frac{xy^{2}}{z^{2}}, Y= \frac{x^{2}y^{2}}{z^{4}}, Z= \frac{xy^{2}}{z^3}
-1 \nonumber 
\end{equation} \\
and x= $-2W_{0}^{'}$ , y = $W_\infty^{'}$ and z =  $W_{0} - W_\infty$. The SPL XC functionals, in turn, is defined as  
\begin{equation}
E_{xc}^{SPL} = (W_0 - W_\infty)[\frac{\sqrt{1+2\chi}-1-\chi}{\chi}] + W_0 
\end{equation} 
with $\chi = \frac{W_{0}^{'}}{  W_\infty - W_0 }$.
}

\subsection{3. The DH functionals}

The DH DFAs have been introduced in \RRef{grimme2007doublehybrid} as an extension of hybrid XC functionals using hybrid-like decomposition in the correlation part of energy expression which reads
\begin{eqnarray}
E^{\rm DH}_{xc} &=&  \xi_1 E^{\rm EXX}_x  + \xi_2 E^{\rm GL2}_c + \xi_{3} E^{\rm DFA}_x + \xi_{4} E^{\rm DFA}_c
\label{eq21}
\end{eqnarray}
where \rev{$\mathbf{\xi} = (\xi_1,\xi_2,\xi_3,\xi_4)$} are the parameters scaling standard semi-local contributions ($E^{\rm DFA}_x$ and  $E^{\rm DFA}_c$), EXX and GL2 terms, accordingly. In most cases, the $\xi_i$ parameters are fixed by empirical fitting to reference data, giving rise to the semi-empirical class of DH functionals. 
The construction was formalized in \Refs{sharkas11}{Rationale_DH} and then in \RRef{ACjustDH1,BApaper,NEil_justDH} through AC formalism. This type of DH is commonly denoted as a non-empirical class. Two limiting cases are important for the DH DFA construction in AC formalism: the $\alpha \to 0$ (given by \Eq{weaklim}) and the full interacting limit at $\alpha \to 1$. In the case of latter, the $W_{xc}^1[\rho]$ is usually approximated by non-local or semilocal formula\cite{levy1985hellmann,levy1985hellmannh2,alipour2015designing} 
derived using the Levy-Perdew scaling relation\cite{levy1985hellmann}
\begin{equation}
W_{xc}^{\alpha}[\rho] =  E^{\rm DFA}_x[\rho] +  2 E^{\rm DFA}_c[\rho_{1/\alpha}]\alpha  + \frac{\partial E^{\rm DFA}_c[\rho_{1/\alpha}]}{\partial \alpha} \alpha^2 
\label{eqn22}
\end{equation}
with $\rho_{1/\alpha}(\R)=\alpha^{-3}\rho(\R/\alpha)$ being the coordinate-scaled\cite{seidl2000density,gorling1993correlation} density which close to the upper limit  ($\alpha =1$) can be replaced by the physical density itself $\rho_{1/\alpha}(\R) \approx \rho(\R)$ leading to a simplified form  $W_{xc}^1[\rho] \approx  E^{\rm DFA}_x[\rho] +  2 E^{\rm DFA}_c[\rho]$. This kind of formalism was successfully applied in the construction of many semi-empirical or non-empirical DH, among which we can recall the PBE0-DH~\cite{bremond11}, PBE0-2~\cite{chai2012seeking}, PBE-QIDH and TPSS-QIDH~\cite{BApaper}, PBE-ACDH\cite{NEil_justDH} or DFA-CIDH\cite{alipour2015designing}, 1DH-LYP\cite{sharkas11} to be the most successful. Additionally, this model has recently been used to support the development of higher-order GL perturbation theory functional\cite{THszs}.

\rev{In this study, we have considered the following semilocal approximations and parametrizations in \Eq{eq21}:
\begin{itemize}
    \item the PBE-QIDH\cite{BApaper} defined using PBE\cite{pbe} semilocal functional ($E^{\rm DFA}_x = E^{\rm PBE}_x$ and $E^{\rm DFA}_c = E^{\rm PBE}_c$) and $\mathbf{\xi} = (0.6933, 0.3066, 0.3333, 0.6667)$
    \item  the B2PLYP \cite{grimme2007doublehybrid} and BL1P functionals\cite{bl1p} defined using B88\cite{B88} ($E^{\rm DFA}_x = E^{\rm B88}_x$) and LYP\cite{LYP} ($E^{\rm DFA}_x = E^{\rm LYP}_x$) functionals together with $\mathbf{\xi} = (0.53, 0.27, 0.47, 0.73)$ and $\mathbf{\xi} = (0.82, 0.6724, 0.18, 0.3276)$ sets of parameters, respectively
    \item the XYG3\cite{XYG3} approximation with $E^{\rm DFA}_x = E^{\rm B88}_x - 0.0664 E^{\rm LDA}_x$, $E^{\rm DFA}_c = E^{\rm LYP}_c$ and $\mathbf{\xi} = (0.8033, 0.3211, 0.2107, 0.6789)$ in \Eq{eq21} \;.
\end{itemize}
}

\clearpage
\bibliography{mp2limit}

\end{document}